\documentclass[]{aa}  

\usepackage{graphicx, natbib, amsmath, longtable}
\bibpunct{(}{)}{;}{a}{}{,}
\def\lsim{\mathrel{\rlap{\lower4pt\hbox{\hskip1pt$\sim$}}
    \raise1pt\hbox{$<$}}}                
\begin{document}
   \title{A VLT/FLAMES survey for massive binaries in Westerlund~1: I. first 
     observations of luminous evolved stars\thanks{This work is 
     based on observations collected at the European Southern Observatory, 
     Paranal Observatory under programme ID ESO 81.D-0324A\dots E.}}

   \author{B.~W.~Ritchie\inst{1,2} \and J.~S.~Clark\inst{1} 
      \and I.~Negueruela\inst{3} \and P.~A.~Crowther\inst{4}}
   \offprints{B.W.~Ritchie, \email{b.ritchie@open.ac.uk}}

   \institute{
        Department of Physics and Astronomy, The Open University, Walton Hall, 
        Milton Keynes MK7 6AA, United Kingdom
   \and
        IBM United Kingdom Laboratories, Hursley Park, Winchester, Hampshire 
        SO21 2JN, United Kindgom
   \and 
        Departamento de F\'{\i}sica, Ingenier\'{\i}a de Sistemas y 
        Teor\'{\i}a de la Se\~{n}al, Universidad de Alicante,
        Apdo. 99, 03080 Alicante, Spain
   \and
        Department of Physics and Astronomy, University of Sheffield, Sheffield
        S3 7RH, United Kingdom
   }

   \date{Accepted ??? Received ???}

   \abstract
   {}
   {Multiwavelength observations of the young massive cluster 
    Westerlund~1 have revealed evidence for a large number of OB supergiant 
    and Wolf-Rayet binaries. However, in most cases these findings are based 
    on the detection of secondary binary characteristics, such as hard X-ray 
    emission and/or non-thermal radio spectra and hence provide little 
    information on binary properties such as mass ratio and orbital period. To 
    overcome this shortcoming we have initiated a long temporal baseline, 
    multi-epoch radial velocity survey that will provide the first direct 
    constraints on these parameters.}
   {VLT/FLAMES+GIRAFFE observations of Wd1 were made on seven epochs from late-June to 
   early-September 2008, covering $\sim$35 confirmed members of Wd1 and 
   $\sim$70 photometrically-selected candidate members. Each target was 
   observed on a minimum of three epochs, with brighter cluster members 
   observed on five (or, in a few cases, seven) occasions. Individual spectra 
   cover the 8484-9001$\text{\AA}$ range, and strong Paschen-series absorption 
   lines are used to measure radial velocity changes in order to identify 
   candidate binary systems for follow-up study.}
   {This study presents first-epoch results from twenty of the most luminous 
    supergiant stars in Wd1. Four new OB supergiant members of Wd1 are 
    identified, while statistically significant radial velocity changes are 
    detected in $\sim$60\% of the targets. W43a is identified as a short-period 
    binary, while W234 and the newly-identified cluster member W3003 are 
    probable binaries and W2a is a strong binary candidate. The cool 
    hypergiants W243 and W265 display photospheric pulsations, while a number 
    of early-mid B supergiants display significant radial velocity changes of 
    $\sim$15--25kms$^{-1}$ that we cannot distinguish between orbital or 
    photospheric motion in our initial short-baseline survey. When combined 
    with existing observations, we find 30\% of our sample to be binary (6/20)
    while additional candidate binaries support a binary fraction amongst Wd1 supergiants 
    in excess of $\sim$40\%, a figure that is likely to increase as further data 
    become available.}
   {}

  \titlerunning{Massive binaries in Wd1}
  \maketitle
%

\section{Introduction}

Despite direct observational confirmation of stars with dynamical masses in 
excess of 80~M$_{\odot}$ (e.g. \citealt{rauw}; \citealt{schnurr}), the production of massive stars
is poorly understood, largely as a result of their intrinsic rarity, apparent 
rapidity of formation and extreme extinction due to veiling by their natal 
envelopes. Moreover, their production is problematic on theoretical grounds, 
primarily because accretion rates must be both extreme in order to build the 
star and yet must overcome the resultant radiation pressure. Hence it is not 
clear if the process is simply a scaled up version of low mass formation or 
whether it proceeds by a different mechanism (e.g. \citealt{bb05} and refs. 
therein). An additional problem is that even if radiation pressure can be 
overwhelmed, sufficient material must be present to accrete from in order to 
yield the stars themselves \citep{davies}.

Nevertheless, several differing scenarios have been advanced to surmount these 
difficulties. Disc mediated accretion has been suggested to  overcome radiation
pressure \citep{yorke} but sufficient material must still be supplied to the 
protostar. One mechanism proposed to accomplish this is competitive accretion 
onto high mass protostars in cluster cores \citep{bb06}. An alternative, which 
also circumvents the constraints imposed by radiation pressure, is that massive
stars form as the result of stellar mergers (e.g. \citealt{b98}, 
\citealt{davies}), though both processes may be relevant \citep{bb05}. 

While direct observations of \textit{in situ} massive star formation are 
technically highly challenging, significant observational constraints on this 
process exist. For instance, the finding that massive stars appear to form in 
clusters (e.g. \citealt{z93}, \citealt{cc}) 
appears difficult to reconcile with turbulent fragmentation of the molecular 
cloud leading to the production of protostellar cores, since the resultant 
clump masses are significantly lower than required to form  massive O stars 
(\citealt{cb}). A second constraint is provided by the observed binary fraction
and properties of massive stars;  competitive accretion predicts their 
formation via accretion onto wide low mass systems leading to massive close 
binaries ($\sim$1~AU; \citealt{bb05}), while  massive star formation via 
stellar mergers leads to a reduction in the number of primordial binaries, and 
hence requires a very high initial fraction. As such, recent observational 
results demonstrating a large binary fraction amongst massive stars have the 
potential to directly test current theories of star formation 
(e.g. \citealt{g01}, \citealt{kim07}, \citealt{sana08}, \citealt{bosch09}, \citealt{clark09a}). 
The properties of these populations, such as the distribution of 
eccentricities, mass ratio and orbital separations place stringent constraints 
on the production rates and channels of both high and low mass X-ray 
binaries - and ultimately double degenerate systems - as well as both Type 
Ib/c and blue Type II supernovae and potentially Gamma Ray Bursters 
(\citealt{kob07} and refs. therein).

The starburst cluster Westerlund~1 (hereafter Wd1; \citealt{w61}) contains a 
rich population of massive, evolved stars including Wolf-Rayets, OB 
supergiants, yellow hypergiants, a luminous blue variable, and red supergiants
\citep{cncg05}. This unique population has been the subject of extensive study 
in recent years at radio \citep{dougherty}, infra-red \citep{crowther06}, 
visual (\citealt{cn02}; \citealt{cncg05}; \citealt{neg09a}) and X-ray 
wavelengths (\citealt{skinner06}; \citealt{clark08}). These studies have shown 
that 17 of the 24 identified Wolf-Rayet stars are binary (\citealt{crowther06}; 
\citealt{clark08}), while photometric \citep{bonanos07} and X-ray 
\citep{clark08} studies also imply a significant binary fraction amongst the 
transitional OB supergiants that are the evolutionary precursors of the 
Wolf-Rayet population. However, the existing studies are largely based on 
secondary binary diagnostics such as the presence of non-thermal radio emission,
hard and/or high luminosity X-ray emission and a significant near-mid IR 
excess, all of which are thought to be observational signatures of Colliding 
Wind  Binaries (CWBs). Moreover, the single epoch observations used to identify 
CWBs provide little information on the nature of the binary, save that the 
secondary is expected to be massive enough to support a 
powerful stellar wind\footnote{Wind-photosphere interaction may also
result in X-ray emission in systems where the secondary lacks a strong
wind (e.g. the O9V + B1--1.5V binary \mbox{\object{CPD~-41$^{\circ}$7742}}; \citealt{sana05}).}. 
Binaries have been directly detected in only a few cases, e.g. the eclipsing 
binaries \object{W13} and \object{W36} \citep{bonanos07} or the double-lined 
spectroscopic binary \object{W10} \citep{neg09a}.

Consequently we have initiated a long temporal baseline, multi-epoch VLT/FLAMES 
spectroscopic radial velocity (RV) survey designed to identify and 
characterise binaries within Wd1. The sensitivity offered is high enough to 
identify both short- and long-period binaries and is sensitive to low mass 
companions within short period systems. This is the first of three
planned papers, describing target selection, data reduction and analysis 
for our survey, and presenting the first results from a subset of the VLT/FLAMES 
dataset containing a sample of luminous, evolved supergiants observed 
on five epochs between 29/06/2008 and 04/09/2008. Paper II will present 
analysis of our full dataset\footnote{Additional data during 2009 is anticipated 
under ESO program ID 383.D-0633.}, consisting of $\sim$105 spectroscopically-
and photometrically-selected targets observed on a minimum of six epochs during
2008 and 2009. A final paper will present follow-up observations and numerical 
modeling of the observed RV variability (\textit{cf.} \citealt{kob07}) aimed at 
placing limits on the close, high-mass binary fraction, the orbital and mass-ratio 
distributions and the permitted formation channels for low- and high-mass X-ray binaries 
within Wd1.

\section{Observations \& data reduction} \label{sec:obs_data}

\begin{figure*}
\begin{center}
\resizebox{\hsize}{!}{\includegraphics{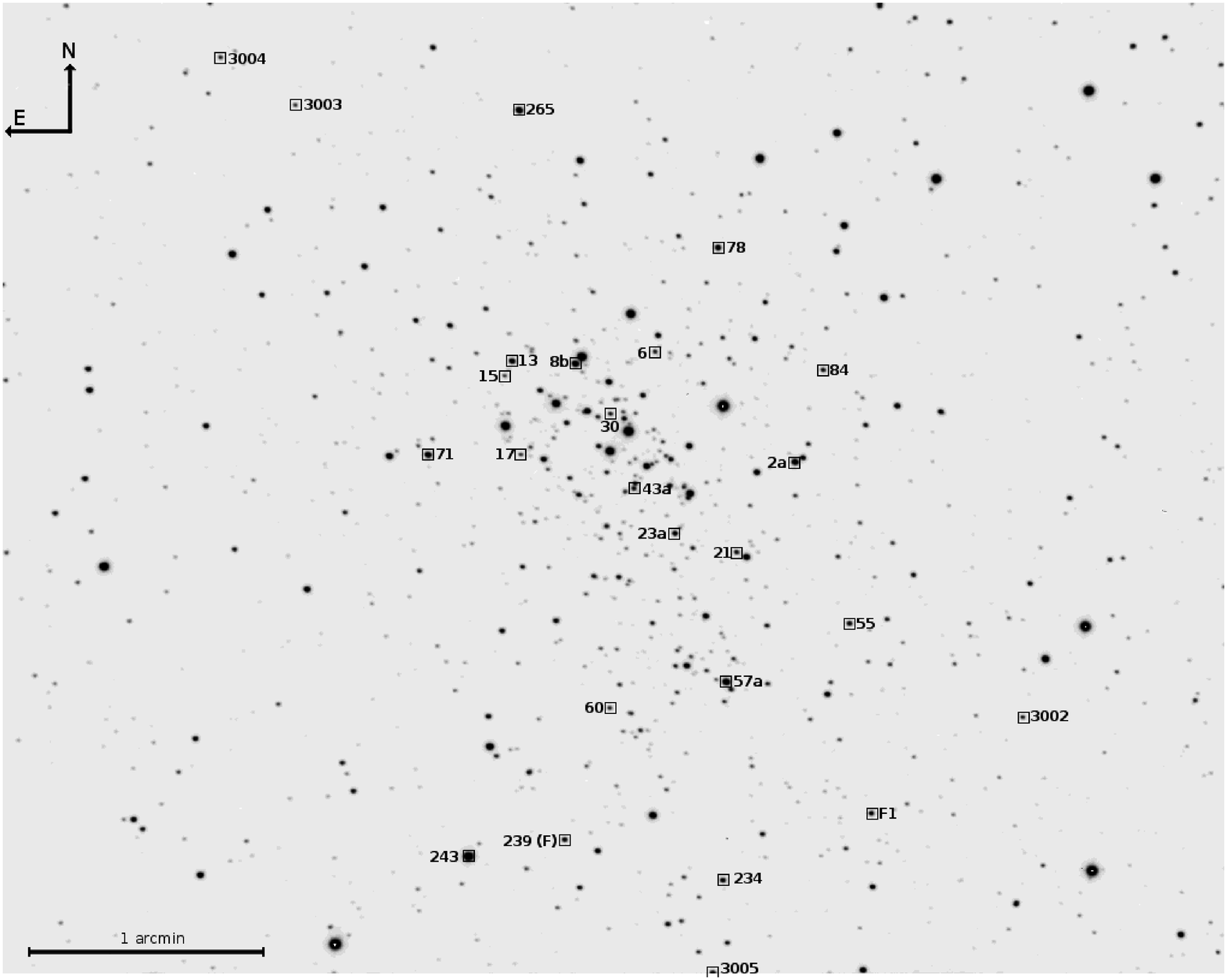}}
\caption{VLT/FORS1 R-band finding chart ($\sim$0.6" seeing) for the targets 
listed in Table~\ref{tab:targets}. The field of view is $\sim$4' 
square, corresponding to $\sim$6pc at an assumed cluster distance of 
$\sim$5kpc \citep{cncg05}. }
\label{fig:finder}
\end{center}
\end{figure*}

\begin{table}
\caption{Summary of observations of the four sets of targets.}
\label{tab:observations}
\begin{center}
\begin{tabular}{l|lc|lc}
Date       & MJD$^{a}$ & $\Delta$t$^{b}$ & Targets & Integration time\\
\hline\hline
&&&&\\
20/06/2008 & 54637.190 & ---             & Faint 1 & 3$\times$895s\\
           & 54637.230 & ---             & Faint 2 & 3$\times$895s\\
29/06/2008 & 54646.185 & ---             & Bright  & 2$\times$600s\\
           & 54646.224 & ---             & Faint 3 & 3$\times$895s\\
\hline
18/07/2008 & 54665.036 & 18.85           & Bright  & 2$\times$600s\\
           & 54665.065 & 18.84           & Faint 3 & 3$\times$895s\\
           & 54665.104 & 27.87           & Faint 2 & 3$\times$895s\\
           & 54665.144 & 27.95           & Faint 1 & 3$\times$895s\\
\hline
24/07/2008 & 54671.134 &  6.10           & Bright  & 2$\times$600s\\
\hline
14/08/2008 & 54692.042 & 20.91           & Bright  & 2$\times$600s\\
           & 54692.073 & 27.01           & Faint 3 & 3$\times$895s\\
17/08/2008 & 54695.094 & 29.99           & Faint 2 & 3$\times$895s\\
           & 54695.133 & 29.99           & Faint 1 & 3$\times$895s\\  
\hline
04/09/2008 & 54713.011 & 20.97           & Bright  & 2$\times$500s\\
\hline 
\end{tabular}
\end{center}
$^{a}$The modified Julian day (MJD) is given at the midpoint of the sequence of
integrations.\\
$^{b}$The time since the target field was last observed.\\
\end{table}

\begin{table*}
\caption{List of targets.}
\label{tab:targets}
\begin{center}
\begin{tabular}{l|c|llccc|l}
ID        & Spectral Type           &  RA (J2000) & Dec (J2000) &R$^{a}$&I$^{a}$& Field$^{b}$ &Notes$^{c}$\\
\hline
\hline
&&&&&&&\\
W2a       & B2Ia$^{e}$              & 16 46 59.71 & -45 50 51.1 & 14.23 & 11.73 & Br.   & A\\
W8b       & B1.5Ia$^{e}$            & 16 47 04.95 & -45 50 26.7 &  --   &  --   & Br.   & \\
W13       & B0.5Ia$^{+}$+OB$^{e}$   & 16 47 06.45 & -45 50 26.0 & 14.63 & 12.06 & Br.   & X, E (9.20 day)\\
W21       & B0.5Ia$^{e}$            & 16 47 01.10 & -45 51 13.6 & 15.56 & 12.74 & Br.   & A\\
W23a      & B1.5Ia$^{e}$            & 16 47 02.57 & -45 51 08.7 & 14.91 & 12.07 & Br.   & A\\
W30a      & O9--B0.5Ia bin$^{f}$    & 16 47 04.11 & -45 50 39.0 & 15.80 & 13.20 & Br.   & X\\
W43a      & B0Ia$^{e}$              & 16 47 03.54 & -45 50 57.3 & 15.22 & 12.26 & Br.   & A\\
W55       & B0Iab$^{g}$             & 16 46 58.40 & -45 51 31.2 & 15.25 & 12.67 & Br./1 & A\\
W60       & B0Iab$^{e}$             & 16 47 04.13 & -45 51 52.1 & 15.96 & 13.28 & Br.   & A\\
W71       & B2.5Ia$^{e}$            & 16 47 08.44 & -45 50 49.3 & 14.06 & 11.16 & Br./1 & A\\
W78       & B1Ia$^{e}$              & 16 47 01.54 & -45 49 57.8 & 14.54 & 12.04 & Br.   & A\\
W84       & O9.5Ib$^{e}$            & 16 46 59.03 & -45 50 28.2 & 15.60 & 13.63 & Br.   & X?\\ 
W234      & O9.5Ib                  & 16 47 01.44 & -45 52 35.0 &  --   &  --   & Br./1 & \\ 
W239 (F)  & WC9d$^{g}$              & 16 47 05.21 & -45 52 25.0 & 15.39 & 12.90 & Br.   & X, A\\
W243      & LBV (A3Ia$^{+}$)$^{h}$  & 16 47 07.55 & -45 52 28.5 &  --   &  --   & Br.   & A\\
W265      & F5Ia$^{+ g}$            & 16 47 06.26 & -45 49 23.7 & 13.62 & 10.54 & Br.   & A\\ 
W373      & B0Iab                   & 16 46 57.71 & -45 53 20.0 & \textit{14.74} & \textit{12.32} & Br.   & A\\
W3002     & B0Iab                   & 16 46 54.24 & -45 51 54.7 & --    & --    & Br.   & \\
W3003     & B0Ib                    & 16 47 11.60 & -45 49 22.4 & \textit{16.21} & \textit{13.31} & Br.   & A\\
W3004     & B0Iab                   & 16 47 13.39 & -45 49 10.5 & \textit{15.96} & \textit{12.99} & Br.   & A\\
W3005     & O9.5Ib                  & 16 47 01.69 & -45 52 57.8 & \textit{15.39} & \textit{12.97} & Br.   & A\\
F1$^{d}$  & M2II-III                & 16 46 57.84 & -45 52 18.5 & --    & --    & Br.   & Field\\
\hline
W6a       & O9--B0.5Ia bin$^{f}$    & 16 47 03.04 & -45 50 23.6 & 15.80 & 13.16 &  1    & X, P (2.20 day)\\
W15       & O9Ib$^{e}$              & 16 47 06.63 & -45 50 29.7 & 16.38 & 13.75 &  2    & X, A\\
W17       & O9Iab$^{e}$             & 16 47 06.25 & -45 50 49.2 & 16.19 & 13.56 &  1    & X\\
W57a      & B4Ia$^{e}$              & 16 47 01.35 & -45 51 45.6 & 13.83 & 11.13 &  1    & A\\
\hline
\end{tabular}
\end{center}
$^{a}$Photometric \textit{R} and \textit{I}-band photometric magnitudes are taken from \cite{cncg05} or \cite{bonanos07} 
where available. Values from \cite{bonanos07} are listed in italics.\\
$^{b}$Targets are listed as \textit{Br.} for members of the \textit{bright} list, 
or 1\dots3 for the \textit{faint} target lists. The stars W6a, W15, W17
and W57a are not included in our main target list, but are discussed in the text.\\
$^{c}$Stars are noted as \textbf{E}clipsing, \textbf{P}eriodic or \textbf{A}periodic variables \citep{bonanos07} 
and/or \textbf{X}-ray sources \citep{clark08}.\\ 
$^{d}$The target F1 is a foreground star, and not a member of Wd1.\\
References: $^{e}$\cite{neg08, neg09a}; $^{f}$\cite{clark08}; $^{g}$\cite{cncg05}; $^{h}$\cite{ritchie09}.\\
\end{table*}

\begin{figure*}
\begin{center}
\resizebox{\hsize}{!}{\includegraphics{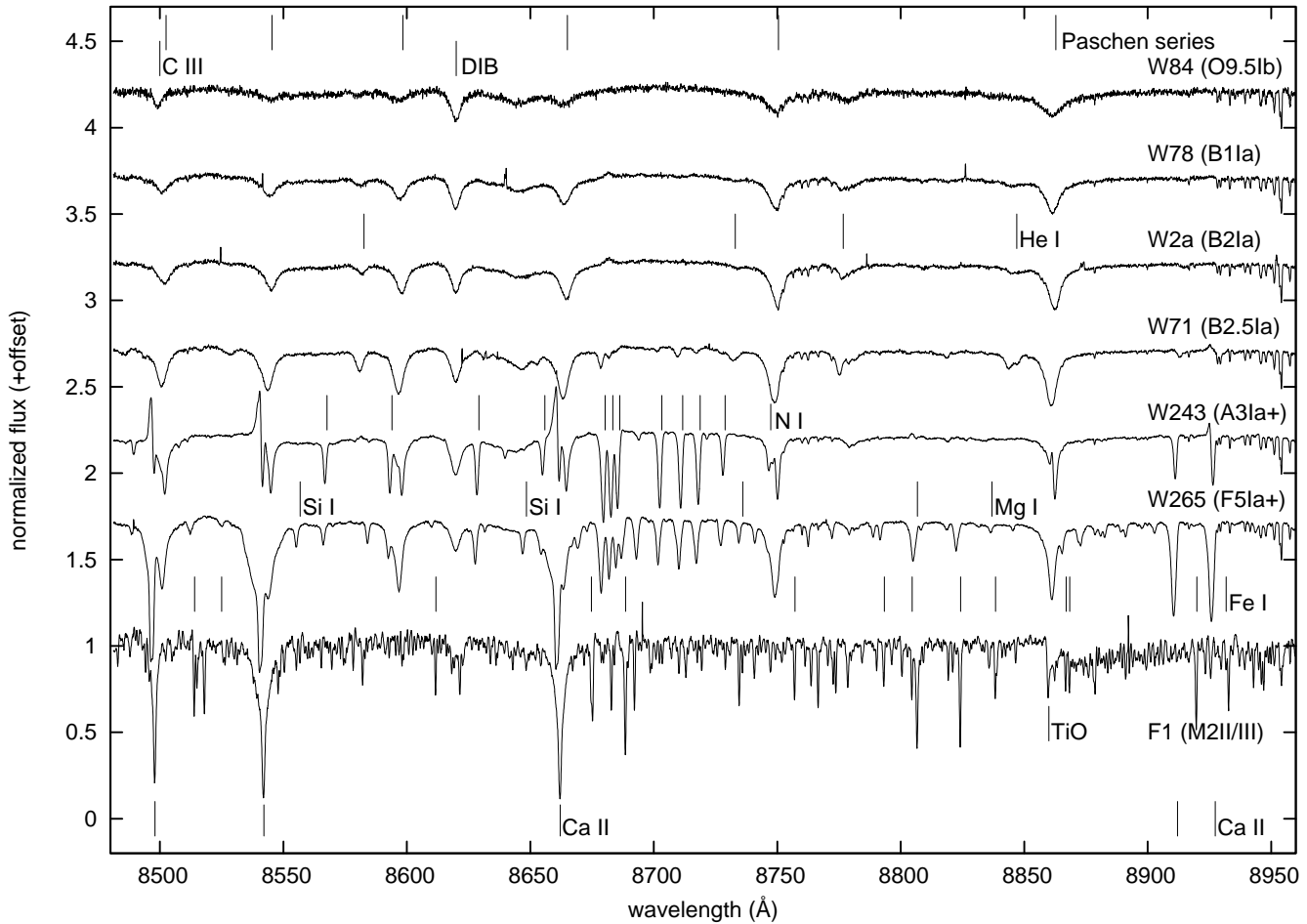}}
\caption{Spectral sequence of selected targets, from the M2II/III field star \object{F1} 
to the O9.5Ib cluster member \object{W84} with the rest wavelengths of the principal 
absorption lines marked. Residual sky lines, as discussed in Section~\ref{sec:obs_data}, are 
visible in some spectra.}
\label{fig:seq}
\end{center}
\end{figure*}

Observations were made using the Fibre Large Array Multi Element Spectrograph 
(FLAMES; \citealt{pasq02}), located on VLT UT2 \textit{Kueyen} at Cerro 
Paranal. FLAMES provides multi-object spectroscopy using the GIRAFFE  medium-high resolution 
and/or UVES high resolution spectrographs, with the red arm of UVES fed by up to eight 
fibres and the MEDUSA mode of GIRAFFE allowing simultaneous observation of up to 132 separate 
objects\footnote{Two GIRAFFE Integral Field Unit modes are also available.}.
Our observations use GIRAFFE in MEDUSA mode, but the high degree of crowding towards the core 
of Wd1 means that we cannot simultaneously allocate fibres to all 
spectroscopically-confirmed cluster members due to limitations imposed by
the physical size of the MEDUSA fibres and the need to avoid collisions. Instead 
we use \textit{FPOSS} (the FLAMES Fibre Positioner Observation Support 
Software) to optimally allocate fibres to targets drawn from four
lists of candidates:
\begin{itemize}
\item Known cluster members catalogued by \cite{w87}. The brightest OB and 
transitional supergiant members of Wd1 are assigned the highest priority for 
fibre allocation, while other catalogued OB I/II stars are assigned moderate 
priority. Yellow hypergiants and the Wolf-Rayet population are assigned low 
priority; the YHGs are presumed too large to allow a close companion, while 
the binary fraction of the Wolf-Rayet population has been examined by 
\cite{crowther06}. 
\item Candidate cluster members with SuSI2 photometry \citep{cncg05}. These 
targets are located in a 5.5$\times$5.5-arcminute field centred just 
north-east of the core of Wd1. These are again assigned moderate priority, with 
increasing likelihood of selection away from the highest-priority OB 
supergiants in the cluster core.
\item Bright stars away from the cluster core with WFI photometry 
\citep{baade99} suggestive of cluster membership. These are assigned low 
priority, but still have a high probability of selection during fibre 
configuration due to the lack of alternative targets away from the cluster 
core. 
\item Objects around the cluster core that are likely to be members based on 
their general colour, but have no existing photometry or spectroscopy. These 
targets are assigned a low priority for fibre configuration and serve as 
backup targets for fibres that cannot be allocated in any other way.
\end{itemize}
Four lists of targets were generated using \textit{FPOSS}. A \textit{bright} 
list, observed on five epochs in 2008, contains 17 of the brightest confirmed 
cluster members and five candidate members located towards the core of Wd1 for 
which no photometry or spectroscopy exists. In addition, three \textit{faint} 
lists, observed on three epochs, contain a further 17 known cluster members, 
58 photometrically-selected candidates (28 with SuSI2 photometry and 30 with 
WFI photometry) and 16 candidates for which no existing observations are 
available. A summary of all the observations is given in Table~\ref{tab:observations}. 
Each target list also includes $\sim$10 fibres that were used to obtain sky spectra. 
Despite being assigned low priority, two Wolf-Rayets, \object{W239}(F/WR77n) and 
\object{W39b}(G/WR77j), were included by the fibre allocation 
process, with \object{W239} included in the \textit{bright} targets list, 
while three targets are duplicated on the \textit{bright} and \textit{faint1} 
lists and are therefore observed seven times in total. With the exception 
of two of these duplicated objects (\object{W71}, B2.5Ia and \object{W55}, B0Iab),
the transitional supergiant \object{W57a} (B4Ia, \textit{faint1}), and the Wolf-Rayet 
\object{W39b}/G (WN7o, \textit{faint2}), the known cluster members in the \textit{faint} lists 
all have spectral classifications in the range O9II--B0Ib \citep{neg09a}, while 
visual inspection of the photometrically-selected targets suggest they are predominantly
reddened late-OI/II stars that are likely \textit{bona
fide} members of Wd1 along with a number of field stars with colours similar to the
reddened supergiants: full spectral classification will be given in Paper II. As many of our 
photometrically-selected targets are uncatalogued by \cite{w87}, we name 
targets that are subsequently confirmed as members of Wd1 as W2$xxx$ if they 
are selected using photometry from \cite{cncg05} and W3$xxx$ otherwise, where 
in both cases the final three digits count upwards from 001. 

Due to the high degree of reddening towards Wd1 
($A_{v}\sim12$; \citealt{cncg05}) the expected signal-to-noise ratio 
at the blue end of the spectrum would be too poor to allow high-precision RV measurements to be 
obtained. Instead, we use the near-IR atmospheric window at 
$\sim8350-9000\text{\AA}$, which covers a number of strong, well defined 
Paschen-series absorption lines in the spectra of OB supergiants 
\citep{cncg05,neg09a}. This region is free from telluric features, and the 
line formation region deep in the stellar photosphere implies the observed 
absorption lines should be free from wind contamination\footnote{To test this 
hypothesis we have calculated non-LTE model atmosphere synthetic spectra for a 
range of parameters, with the synthetic spectra showing no evidence for wind 
contamination, even with mass loss rates an order of magnitude above expected 
values.}. The GIRAFFE spectrograph was therefore used with 
setup HR21 covering the 8484-9001$\text{\AA}$ range with $R = \lambda/\Delta\lambda\sim 16200$.
Two 600s integrations were used for the bright targets, while three 895s
integrations were used for the faint targets. The FLAMES data were 
bias-subtracted, flat-fielded to correct for pixel-to-pixel variation and
fibre transmission differences, 
and wavelength-calibrated using version 2.5.3 of the FLAMES-GIRAFFE 
pipeline\footnote{http://www.eso.org/sci/data-processing/software/pipelines/} 
with version 4.1.0 of the Common Pipeline Library (CPL),
with individual spectra extracted from the final pipeline frames using the 
IRAF\footnote{IRAF is distributed by the National Optical Astronomy 
Observatories, which are operated by the Association of Universities for 
Research in Astronomy, Inc., under cooperative agreement with the National 
Science Foundation.} task \textit{onedspec}. The signal-to-noise ratio of
our co-added spectra varies with the $I$-band luminosity of the target, ranging 
from $S$/$N$$>$200 for the very luminous transitional hypergiants \object{W243} 
and \object{W265} to $\sim$65--75 amongst the least luminous O9.5Ib targets in
our \textit{bright} sample. A master sky spectrum was created 
from individual fibre sky spectra, and this was subtracted from stellar spectra
using the \textit{skysub} task. Notable fibre-to-fibre variations in the sky 
spectra are apparent across the FLAMES field, and as a result the removal 
of sky lines from stellar spectra is frequently imperfect (see also discussion
in \citealt{evans05}). As can be seen from Figure~\ref{fig:seq}, residual sky 
lines are not generally significant, but in a few cases unsubtracted sky lines 
overlapped absorption lines used for RV measurement and had to be removed by 
manually interpolating between levels on either side of the emission line. Finally, 
spectra were corrected for heliocentric velocity using \textit{dopcor} and normalized using the 
\textit{continuum} task. A serendipitous check for zero-point errors in our 
data is provided by a strong, well-defined diffuse interstellar band at 
$\sim$8620$\text{\AA}$ \citep{munari00}. As this feature is unassociated with 
Wd1 it displays a constant profile and a RV that varies by less than 
$\sim$1kms$^{-1}$ in all spectra, implying that there are no systematic shifts 
in line centre or profile between epochs.

In this work only the \textit{bright} target list is examined; a list of 
targets is given in Table~\ref{tab:targets} and a finder chart is plotted in 
Figure~\ref{fig:finder}\footnote{\object{W373} lies outside the field of 
view of Figure~\ref{fig:finder} and is located to the south of Wd1 16~arcseconds 
east of the bright field star \object{HD151018}.}. A representative spectral sequence, extending from
the earliest objects in our \textit{bright} sample (O9.5Ib, e.g. \object{W84} and
\object{W234}) is plotted in Figure~\ref{fig:seq} with the Paschen-series and He~I lines labeled, 
along with the DIB at $\sim$8620$\text{\AA}$, the C~III~$\lambda$8500 line that blends with the 
Pa-16 line in stars of B0.5I and earlier, and the neutral and singly-ionized metal lines that become 
prominent in the cooler stars. Radial velocities were 
determined from the strong Paschen-series and (when available) He~I absorption 
lines, with the Pa-16~$\lambda$8502 line excluded in stars earlier than B1I due to 
blending with C~III~$\lambda$8500 (this effect can be clearly seen in the bluewards 
shift in the Pa-16/C~III blend in the O9.5Ib spectrum plotted in Figure~\ref{fig:seq}). 
In the case of the cool hypergiants \object{W243} and \object{W265}, the 
Paschen-series lines are frequently blended with adjacent Ca~II and N~I lines, 
and instead nine strong, unblended N~I absorption lines from high-excitation 
multiplets ($\chi_{\text{low}}\sim$10.2eV) were used.  
Measurements were made in a similar manner to \cite{bosch09}, by using the 
IRAF \textit{ngaussfit} task within the \textit{stsdas} package to fit 
Gaussian profiles to the absorption lines, with the measured stellar radial 
velocity at each epoch an error-weighted average of the individual absorption 
lines. Rest wavelengths for RV measurement are taken from the NIST Atomic Spectra
database\footnote{http://physics.nist.gov/PhysRefData/ASD/lines\_form.html}.
Errors are $\lsim$4kms$^{-1}$ unless noted.  

\section{Results}\label{sec:results}

The \textit{bright} target list contains 16 stars that are known members of 
Wd1 with previously published spectral classifications 
\citep{cncg05, clark08, neg08}. The previously-unclassified target 
\object{W373}, located to the south of Wd1 16~arcseconds from the field star 
\object{HD151018}, is classified as B0Iab based on the strength the 
Paschen-series and He~I absorption lines, with a spectrum very similar to 
\object{W60} (also B0Iab; \citealt{neg09a}). Four of the five candidate targets
are confirmed as members of Wd1. \object{W3002} and \object{W3004} display 
very similar spectra to \object{W373}, and are again classified as B0Iab.
\object{W3005} is slightly earlier, displaying a similar spectrum to 
\object{W84} and \object{W234} (O9.5Ib; \citealt{neg09a}), while \object{W3003}
displays slightly broader Paschen-series lines than the B0Iab stars and is 
classified as B0Ib. Finally, although the fifth candidate target (F1) has a similar 
colour to the highly-reddened OB supergiants in Wd1, the lack of a well-defined DIB 
at 8620$\text{\AA}$ and the strength of the TiO $\lambda$8860 bandhead 
(\citealt{ramsey}) show it to be a foreground early-M giant and not a cluster 
member.

Table~\ref{tab:radial} lists the measured radial velocities for the target
stars. In the case of \object{W30a}, the I-band spectrum 
shows only very broad, weak Paschen-series absorption 
lines that preclude accurate measurement; we return to this in 
Section~\ref{sec:discussion}. Selected targets are examined further in the
following sections. Although the detection of binaries in Wd1 is the long-term 
goal of our project, at this stage we are also interested in characterising 
possible photospheric sources of variations in RV. Both sources
of variability are therefore discussed.

\subsection{The eclipsing binary W13}
\label{sec:w13}

\begin{figure}
\begin{center}
\resizebox{\hsize}{!}{\includegraphics{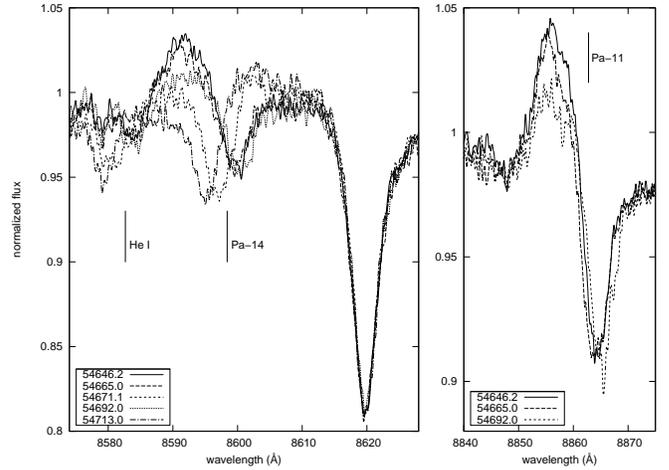}}
\caption{Pa-14~$\lambda$8598 (all five epochs, left panel) and Pa-11~$\lambda$8862
(three epochs at near-identical phase, right panel) in the spectrum of 
\object{W13}. The absorption feature at $\sim$8620$\text{\AA}$ is a DIB, and rest wavelengths are marked. In 
this and subsequent figures the key gives the modified Julian date on which 
each spectrum was obtained.}
\label{fig:w13dib}
\end{center}
\end{figure}

\cite{bonanos07} report \object{W13} to be a contact eclipsing binary with 
a 9.2-day orbital period, and it is also listed as an X-ray source with 
L$_{x}\sim$10$^{32}$ergs$^{-1}$ by \cite{clark08}.  
\cite{neg09a} classify \object{W13} as a B0.5Ia$^{+}$+OB binary, with one 
component being an emission-line object displaying strong H$\alpha$ emission
alongside C~II~$\lambda\lambda$6578, 6583 and He~I~$\lambda\lambda$6678, 7065 
lines in VLT/FLAMES LR6-mode spectra \citep{clark09}. 
By chance, three of our observations are separated by almost integer multiples 
of this orbital period: the second spectrum was obtained 2.05 orbital periods 
after the first, while the fourth spectrum was taken after 4.98 orbital periods 
had elapsed. The other two are out of phase, taken after 2.71 and 7.26 orbital 
periods. We would therefore expect to see absorption lines at almost the same RV 
in three spectra with two discrepant, and this can be seen in Figure~\ref{fig:w13dib}. 
An emission component is seen in all Paschen-series lines, weakening 
noticeably between the first and fourth spectra. Weak He~I absorption features 
are also seen, with He~I~$\lambda\lambda$8583, 8777 \citep{vh72} most prominent.
The radial velocities of the He~I absorption lines closely match the absorption 
components of the Paschen-series lines, and although no obvious He~I 
emission is detected in our I-band spectra the expected strength is low. Radial
velocity measurements and the virtual absence of an emission line component at phase
$\sim$0.7 (corresponding to the 24/07/2008, MJD=54671.1 spectrum) when the 
emission-line component is near (or at) eclipse show \object{W13} to be a spectroscopic 
binary, with the emission component originating in a B0.5I$^{+}$/WNVL primary and the absorption 
component in the secondary. The weakness of the Paschen-series absorption lines 
suggests a $\sim$O8-9I classification for the secondary, but this is inconsistent 
with the observed He~I absorption lines which imply a later spectral type, and it is 
likely that there is significant infilling of the Paschen-series lines from the 
wind of the emission-line primary. \object{W13} therefore appears to be an 
immediate precursor to the binary-rich Wolf-Rayet population in Wd1 \citep{crowther06}.
Radial velocity measurements and orbital parameters for \object{W13} will appear in 
a forthcoming paper. 

\subsection{The Wolf-Rayet W239 (F/WR77n)}
\label{sec:w239}

\begin{figure}
\begin{center}
\resizebox{\hsize}{!}{\includegraphics{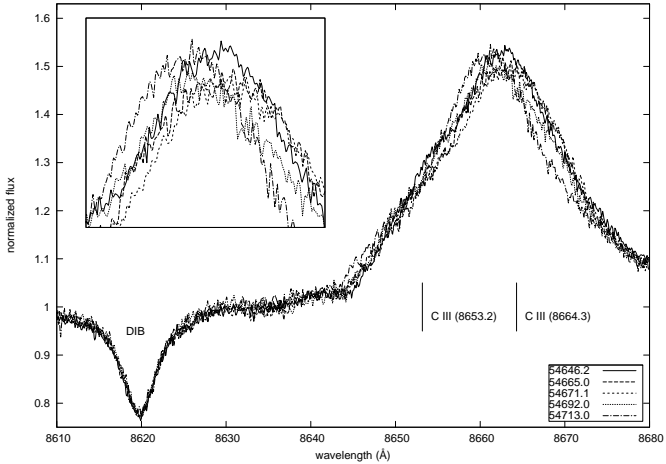}}
\caption{C~III~$\lambda$8664 and the adjacent $\sim$8620$\text{\AA}$ DIB
in the Wolf-Rayet \object{W239}/F. A secondary component at $\lambda$8653 responsible 
for the extended wing bluewards of the main emission peak is also marked.}
\label{fig:w239dib}
\end{center}
\end{figure}

The WC9d star \object{W239} (F/WR77n in the nomenclatures of \citealt{cn02} and 
\citealt{vdh06} respectively) has previously been highlighted as a likely binary
due to its strong near-IR excess and hard X-ray emission (\citealt{crowther06}; \citealt{clark08}), 
while \cite{dougherty} report radio emission that is also consistent with a CWB. 
Our spectra are dominated by two strong C~III$\lambda\lambda$8500,8664 
emission lines, with weak C~IV$\lambda$8856 emission also apparent; the 
C~III~$\lambda$8664 line and nearby DIB at $\sim$8620$\text{\AA}$ are 
plotted in Figure~\ref{fig:w239dib}. Significant radial velocity changes are 
apparent in all three emission lines, although the measured values of the two C~III
lines generally differ by $\sim$20--40kms$^{-1}$, a result of excess emission on the
redwards side of the C~III~$\lambda$8500 line shifting the profile fit relative
to C~III~$\lambda$8664 and also possibly the influence of a weak secondary 
component at 8653$\text{\AA}$ shifting the C~III~$\lambda$8664 fit bluewards. 
Measurement is also complicated by the emission line profiles, which are 
sometimes skewed or weakly double-peaked. However, over the course of our 
observations RVs measured from the C~III~$\lambda$8664 line span 
a range of -82kms$^{-1}$ to +14kms$^{-1}$, with the C~III~$\lambda$8500 line giving a 
very similar overall range that is redshifted by $\sim$25kms$^{-1}$ (errors 
estimated at $\lsim$10kms$^{-1}$ in both cases), with a large delta of $\sim$65kms$^{-1}$
between the 24/07/2008 (MJD=54671.1) and 14/08/2008 (MJD=54692.0) spectra. Although
WRs typically display low-level variability in emission lines due to clumping
in the wind, the RV changes seen here are strongly suggestive of binarity. 
Dust-forming WC stars are generally considered to be binary systems with an 
OB companion (e.g. \citealt{tuthill99}), and the observed RVs are consistent 
with a $\sim$15M$_{\odot}$ WR primary \citep{crowther06} and more massive OB secondary 
in a short-period orbit viewed at moderate inclination (unlike B/WR77o, eclipses 
are not visible in W239; \citealt{bonanos07}). The limited sampling of our five epochs 
precludes estimation of an orbital period, but our long-term dataset 
will allow us to constrain the orbital parameters of \object{W239}. 

\subsection{The spectroscopic (candidate) binaries W2a, W43a, W234 and W3003}

\begin{figure}
\begin{center}
\resizebox{\hsize}{!}{\includegraphics{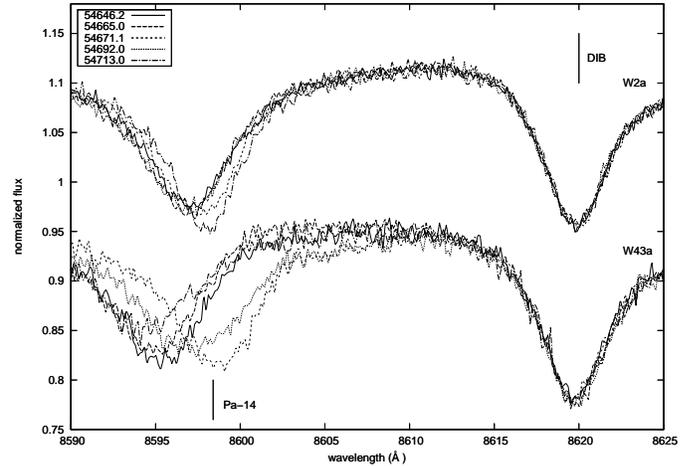}}
\caption{Pa-14~$\lambda$8598 and adjacent $\sim$8620$\text{\AA}$ DIB
in the spectrum of \object{W2a} (top) and \object{W43a} (bottom). The Pa-14 rest 
wavelength marked. \object{W234} and \object{W3003}, the two other 
candidate binaries discussed in the text, are omitted for clarity and are 
intermediate between the two plotted spectra.}
\label{fig:w2a_w43a_dib}
\end{center}
\end{figure}

Figure~\ref{fig:w2a_w43a_dib} plots the Pa-14~$\lambda$8598 line and the 
adjacent DIB at 8620$\text{\AA}$  for \object{W2a} and \object{W43a}, with 
both objects
displaying notable RV variations in the photospheric absorption lines.
\object{W43a} is unambiguously a short-period binary, with the measured radial 
velocities spanning a range of $\sim$140kms$^{-1}$ and including a delta of 
$\sim$131kms$^{-1}$ in the space of six days between 18/07/2008 (MJD=54665.0) 
and 24/07/2008 (MJD=54671.1). No eclipses are reported by \cite{bonanos07}, 
but the relatively high radial velocities suggest that the system must be 
near the \textit{sin~i}$\sim$0.7 limit where eclipses would become visible. 
The $\sim$35M$_{\odot}$ initial mass of the B0Ia primary requires a massive 
companion to produce the observed RV changes, and although there 
is not direct spectroscopic evidence for this secondary, changes in the He~I 
line profiles are tentatively suggestive of an OB companion. However, we note
that changes in the Paschen-series absorption line strengths are also apparent
in our spectra, and it seems likely that we are also observing photospheric 
variations superimposed on the changes in line centre due to orbital motion. 
This may explain changes in the He~I profile, and follow-up observations at 
other wavelengths may be required to identify the companion. Nevertheless, 
\object{W43a} represents an encouraging prospect for accurate parameter 
estimation as further data become available. 

\object{W234} is duplicated on the \textit{bright} and \textit{faint1} lists
and therefore observed on seven occasions. The overall range of observed RVs
is far smaller than \object{W43a} ($\sim$36kms$^{-1}$), but relatively rapid 
RV changes are again present with spectra from 18/07/2008 (MJD=54665.0) and 
24/07/2008 (MJD=54671.1) showing a delta of $\sim$35kms$^{-1}$. This is large 
for bulk photospheric motions in the Paschen-series line forming region, and 
it is more likely that \object{W234} is a similar short-period binary to 
\object{W43a} but viewed at a less favourable angle for RV measurement. 
\object{W2a} displays RVs ranging from -56.2kms$^{-1}$ to 
-18.6kms$^{-1}$ over the course of our observations, including a change of 
$\sim$22kms$^{-1}$ within the six days separating the 18/07/2008 (MJD=54665.0) 
and 24/07/2008 (MJD=54671.1) spectra, and a change of $\sim$36kms$^{-1}$ between
the final two spectra. The variability on short timescales is less pronounced than 
in \object{W234}, but the amplitude of the RV changes is still large relative to 
the majority of our sample. However, we cannot categorically rule out a photospheric origin for 
these variations and classify \object{W234} as a \textit{probable} binary
based on the magnitude of its short-term RV changes and 
\object{W2a} as a \textit{candidate} binary pending further observation. 

Finally, the newly-confirmed cluster member \object{W3003} displays RVs
spanning $\sim$50kms$^{-1}$ over the course of our observations. Unlike the 
other short-period (candidate) binaries identified here, \object{W3003} shows 
no rapid deltas in RV but increases from -4.4kms$^{-1}$ in the first spectrum 
(29/06/2008, MJD=54646.2) to -54.6kms$^{-1}$ in the third (24/07/2008, 
MJD=54671.1) before turning around and decreasing to -10.6kms$^{-1}$ in the 
final spectrum (04/09/2008, MJD=54713.0). Like \object{W234}, additional 
observations are required to confirm the binary nature of the system, but the 
pattern of RV changes appears best explained by a limited sampling of a 
longer-period orbit viewed at moderate inclination. We therefore classify 
\object{W3003} as a second \textit{probable} binary.  

\subsection{The pulsating hypergiants W243 and W265}
\label{sec:hypergiants}

\begin{figure}
\begin{center}
\resizebox{\hsize}{!}{\includegraphics{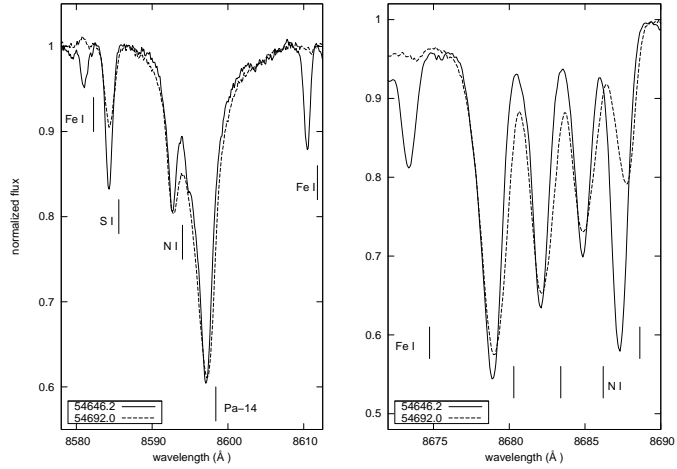}}
\caption{Spectral variability in W265. The left panel plots the region around 
the Pa-14~$\lambda$8598 line, while the right panel plots the region around 
the N~I triplet at $\lambda$8680\dots6. Rest wavelengths are indicated.}
\label{fig:w265p1}
\end{center}
\end{figure}

\begin{figure}
\begin{center}
\resizebox{\hsize}{!}{\includegraphics[angle=270]{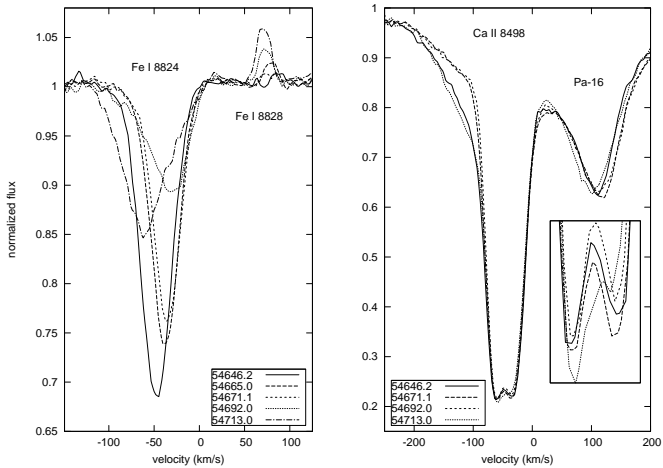}}
\caption{Spectral variability in W265. The left panel plots Fe~I~$\lambda$8824 
(multiplet 60) and $\lambda$8828 (multiplet 1269), with velocities relative to 
the Fe~I~$\lambda$8824 rest wavelength. The right panel plots the 
Ca~II~$\lambda$8498/Pa-16~$\lambda$8503 blend, with velocities relative to the 
Ca~II~$\lambda$8498 rest wavelength.}
\label{fig:w265p2}
\end{center}
\end{figure}

\begin{figure}
\begin{center}
\resizebox{\hsize}{!}{\includegraphics{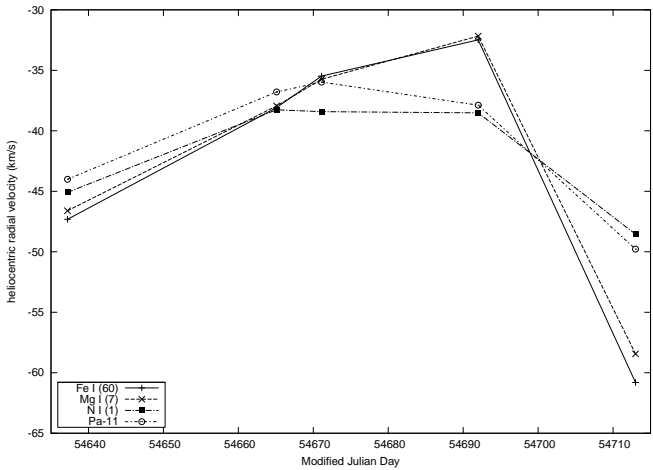}}
\caption{Heliocentric radial velocity changes in low- and high-excitation
absorption lines in \object{W265}. Plotted are Fe~I (multiplet~60, 
$\chi_{\text{low}}\sim$2.2eV), Mg~I (multiplet~7, $\chi_{\text{low}}\sim$4.3eV),
N~I (multiplet~1, $\chi_{\text{low}}\sim$10.2eV) and the Paschen-11 line
($\chi_{\text{low}}$=12.04eV). Errors are $\sim\pm$2kms$^{-1}$.}
\label{fig:w265vel}
\end{center}
\end{figure}

\object{W243} (LBV $\sim$A3Ia$^{+}$; \citealt{ritchie09}) and \object{W265} 
(F5Ia$^{+}$; \citealt{cncg05}) represent two of the population of cool 
hypergiants in Wd1. Turning first to the yellow hypergiant \object{W265},
the VLT/FLAMES spectra reveal pronounced spectral variability in the 
neutral metal lines, and in particular in low-excitation multiplets of Fe~I,
Mg~I and Si~I. This can be seen in Figure~\ref{fig:w265p1}, which plots the
regions around the Pa-14~$\lambda$8598 line and the N~I triplet at 
$\sim$8683$\text{\AA}$ on two epochs, separated by $\sim$46 days, that correspond 
approximately to apparent minimum and maximum $T_{\text{eff}}$ (29/06/2008, MJD=54646.2 and 14/08/2008, MJD=54692.0 
respectively). The first spectrum displays strong Fe~I lines from multiplets 60 
(a$^{5}$P-z$^{5}$P$^{\circ}$, $\chi_{\text{low}}\sim$2.2eV), 339
(b$^{3}$P-z$^{3}$P$^{\circ}$, $\chi_{\text{low}}\sim$2.8eV) and 401 
(b$^{3}$G-z$^{3}$G$^{\circ}$, $\chi_{\text{low}}\sim$3eV), but in the later 
spectrum the multiplet 60 line ($\lambda$8689) has weakened substantially 
while the
multiplet 339 lines ($\lambda\lambda$8611, 8675) are undetectable and 
the multiplet 401 line ($\lambda$8582) is weakly in emission. This pattern
continues in the left panel of Figure~\ref{fig:w265p2}\footnote{Note that for clarity only four epochs of 
data are shown in the right panel, with the omitted 17/08/2008 (MJD=54665.0) 
profile very similar to the plotted profile from 24/08/2008 (MJD=54671.1).}, which plots the 
adjacent Fe~I~$\lambda$8824 (multiplet 60) and $\lambda$8828 (multiplet 1269,
x$^{5}$D$^{\circ}$-e$^{3}$D, $\chi_{\text{low}}\sim$4.9eV) lines over all five 
VLT/FLAMES epochs. Once again, the multiplet 60 line is seen to vary 
significantly in both strength and profile, while the multiplet 1269 line 
strengthens in emission. 
In contrast, the N~I, Ca~II and Pa-16 lines visible in Figures~\ref{fig:w265p1}
and~\ref{fig:w265p2} display much less variability. The Fe~I lines 
($\chi_{\text{low}}\sim$2-3eV) and Mg~I lines ($\chi_{\text{low}}\sim$4eV) 
form in the upper photosphere, while the N~I ($\chi_{\text{low}}\sim$10eV) and
Paschen-series form at lower levels. Notable differences between the two 
layers are apparent in Figure~\ref{fig:w265vel}, which plots the measured 
radial velocities of two low-excitation and two high-excitation lines over the
course of our observations. The outer photosphere, traced by low-excitation 
Fe~I and Mg~I lines, displays a pronounced infall followed by rapid expansion 
in the final epoch. The observed minima in the strengths of the Fe~I absorption
lines corresponds to maximum infall (and presumably maximum $T_{\text{eff}}$), 
with these lines subsequently broadening due to the rapid expansion in the 
final epoch. In contrast the lower layers of the photosphere are more stable, 
and although all species display similar radial velocities during the initial 
infall phase, the high-excitation lines show contraction ending earlier with a 
less pronounced expansion in the final epoch. 

The behaviour of the Fe~I lines is remarkably similar to the well-studied
YHG \object{$\rho$ Cassiopeia} \citep{lobel03}, with the multiplet 60 line in 
Figure~\ref{fig:w265p2} developing a similar triangular profile with excess
absorption in the blue wing\footnote{The excess blue-wing absorption forms
in a cool, optically-thick expanding wind at smaller optical depth 
\citep{lobel03}.} to that reported by \cite{lobel98} for the low-excitation 
Fe~I~$\lambda$5572.8 line. Our limited spectral coverage prevents us from 
examining other low-excitation absorption lines used by \cite{lobel98} to 
interpret the spectrum of \object{$\rho$ Cas}, but, given the similarities, 
it is likely the Fe~I lines share a common origin in a (non-radially) 
pulsating photosphere. 

Finally, we note the weak emission evident in the core of the strong 
Ca~II~$\lambda$8498 line in Figure~\ref{fig:w265p2}; a similar effect
is seen in the core of the other Ca~II $\lambda\lambda$8542,8662 multiplet-2 
lines, but in these cases infilling of the line centre redwards of the
emission peak is seen, leading to an asymmetric core profile. 
The 4$^{2}$P$^{\circ}$ ($\chi_{\text{low}}\sim$1.7eV) lower level of 
the near-IR Ca~II triplet is fed by the Ca~II-K and~-H lines in the near-UV, 
and core emission in the Fraunhoffer lines is an indicator of chromospheric 
activity in cool stars. Such cores are reported in 
\object{$\alpha$ Orionis} but are not observed in the near-IR triplet 
\citep{ld00}, implying a higher chromospheric temperature in \object{W265}.
Observations of \object{$\rho$ Cas} show that core emission in the 
chromospheric Ca~II-H and~K lines is not constant \citep{lobel03}, 
implying that a permanent chromosphere is not present, and this may also be the 
case for \object{W265}; no apparent chromospheric X-ray emission was detected 
\cite[][section 4.2.3]{clark08}, suggesting that the chromospheric 
activity implied by the Ca~II lines may also be transitory. Further 
observation and modeling of \object{W265} is required to determine
the origin of these features. \cite{lobel03} report line-splitting in 
low-excitation ($\chi_{\text{low}}\le$1.5eV) Fe~I multiplets and other 
low-excitation metals. However, our limited spectral coverage precludes 
examination of the multiplets with in which splitting is reported. 
\cite{sgt61} lists Fe~I multiplet 60 as containing double lines in 
\object{$\rho$ Cas}, but no evidence of core doubling in this multiplet 
is seen in our spectra of \object{W265}. 

\begin{figure}
\begin{center}
\resizebox{\hsize}{!}{\includegraphics{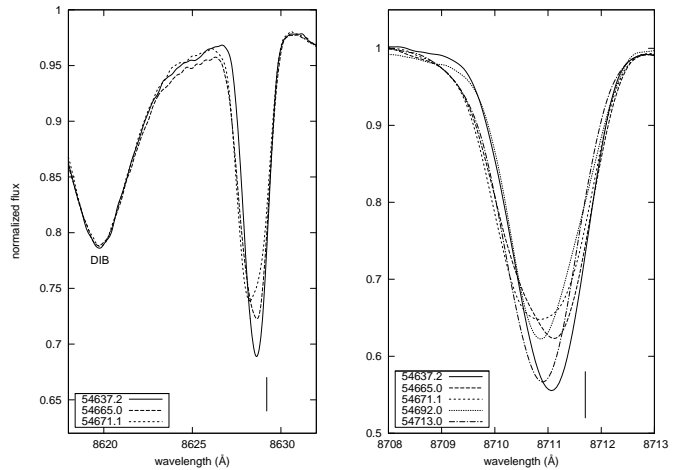}}
\caption{Spectral variability in the LBV W243. The left panel plots 
N~I~$\lambda$8629 line and adjacent $\sim$8620$\text{\AA}$ DIB from three 
epochs (with the other two omitted for clarity) while the right panel plots 
the N~I~$\lambda$8712 line from all five epochs. Rest wavelengths are marked.}
\label{fig:w243}
\end{center}
\end{figure}

Significant pulsational variability is also observed in \object{W243}. The 
highly-variable Fe~I lines observed in \object{W265} are not present in the 
hotter LBV (A3Ia$^{+}$), which displays a complex Fe~II spectrum along with 
absorption lines from other neutral and singly-ionized metals and an 
emission-line spectrum containing Balmer- and Paschen-series lines, He~I 
and Ca~II. This emission-line spectrum is a result of a hot companion star 
ionizing the wind of the LBV primary \citep{ritchie09}, but this is not 
directly reflected in the cool-state absorption lines which display radial 
velocities in a narrow range of -45kms$^{-1}$ to -15kms$^{-1}$ across ten epochs of 
VLT/UVES and VLT/FLAMES data. These lines display RV variations on timescales 
of days, but - if they were due to orbital motion - we would expect a much 
wider range of radial velocities than observed unless, by chance, we are 
observing the system from almost directly above the orbital plane. The 
variations are most clearly seen in the Si~II $\lambda\lambda$6349, 6371 
doublet (obtained using VLT/UVES with cross-disperser \#3, R$\sim$40,000) 
and in the near-IR N~I lines; an example of the latter is shown in 
Figure~\ref{fig:w243}. In addition, VLT/UVES observations of the N~I lines 
reveal the formation of excess bluewards absorption similar to that seen in 
the Fe~I multiplet 60 line in \object{W265} at a time when Fe~II emission lines 
develop pronounced P-cygni profiles indicating enhanced mass loss. 
\object{W243} therefore appears to be undergoing very similar pulsational 
mass-loss to \object{W265}, albeit at higher $T_{\text{eff}}$; we examine the 
spectrum of \object{W243} in detail in \cite{ritchie09}.  

\subsection{The early-B supergiants}
\label{sec:earlyB}

\begin{figure}
\begin{center}
\resizebox{\hsize}{!}{\includegraphics{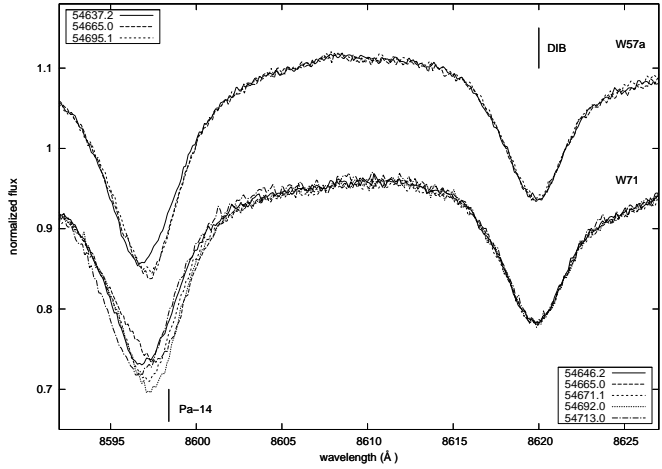}}
\caption{The Pa-14~$\lambda$8598 line and adjacent $\sim$8620$\text{\AA}$ DIB
in the spectrum of \object{W57a} (top) and \object{W71} (bottom). The rest 
wavelength of Pa-14 is marked.}
\label{fig:w57a_w71_dib}
\end{center}
\end{figure}

\begin{figure}
\begin{center}
\resizebox{\hsize}{!}{\includegraphics{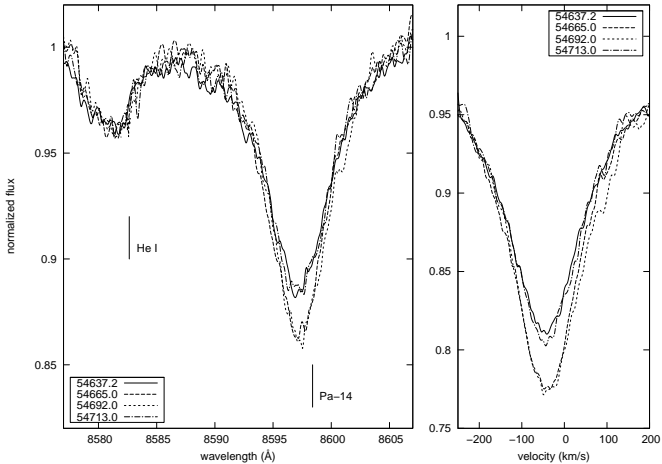}}
\caption{Variations in the Paschen-series absorption lines in \object{W78}. The
left panel plots the Pa-14~$\lambda$8598 line and adjacent He~I~$\lambda$8583 
line, with rest wavelengths marked, while the right panel plots the 
Pa-11~$\lambda$8862 line with velocities relative to its rest wavelength.}
\label{fig:w78}
\end{center}
\end{figure}

Statistically-significant RV changes are also apparent in a number of 
the early-B supergiants, including \object{W8b}, \object{W21}, \object{W23a}, 
\object{W71} and \object{W3004}. Of these, \object{W71} (B2.5Ia; 
\citealt{neg09a}) is one of the latest of the continuous sequence of OB 
supergiants in Wd1, displaying strong Paschen-series and He~I absorption lines 
along with weak N~I lines indicative of its cooler state. \object{W8b} and 
\object{W23a} are slightly earlier, displaying no apparent N~I lines. Finally, 
\object{W21} and \object{W3004} are the earliest of this group, with spectral 
types B0.5Ia and B0Iab respectively. With the exception of \object{W8b} all 
are listed as aperiodic variables by \cite{bonanos07}, and none are 
significant detections at X-ray wavelengths \citep{clark08}. 

\begin{figure}
\begin{center}
\resizebox{\hsize}{!}{\includegraphics{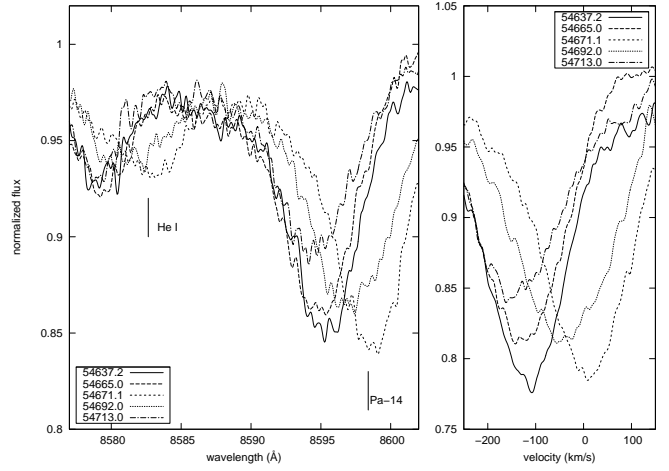}}
\caption{As Figure~\ref{fig:w78}, but showing variations in the Paschen-series 
absorption lines in the short-period binary \object{W43a}.}
\label{fig:w43a}
\end{center}
\end{figure}

In all cases, the observed RV variations lie in a narrow range of 
$\sim$15--25kms$^{-1}$ with no rapid variations between epochs. Although this 
is compatible with longer-period orbital motion, the radial 
velocity changes in the Paschen-series lines are accompanied by changes in 
absorption line strength and profile, and it is more likely that the origin of the RV 
changes is photospheric. This is best illustrated by \object{W71} and the 
cooler transitional supergiant \object{W57a} (B4Ia; \citealt{neg09a}) which 
is included in the \textit{faint1} list and was observed on three occasions; 
both are plotted in Figure~\ref{fig:w57a_w71_dib}. As well as variations in 
the Paschen-series lines, both objects also show pronounced changes in the 
strength of the N~I~$\lambda$8680 triplet over the course of our observations, 
a sensitive indicator of temperature that strengthens rapidly from B2Ia to B4Ia 
\citep{neg09a} and is therefore strongly suggestive of photospheric changes. 
In addition, \object{W78} (B1Ia) does not display statistically significant RV 
changes ($\lsim$10kms$^{-1}$ during the course of our observations) but 
nevertheless displays notable changes in the strength of the Paschen-series 
absorption lines. These changes, plotted in Figure~\ref{fig:w78}, are not 
apparent in the adjacent He~I line which again suggests a photospheric origin. 
Radial velocity or line profile changes are therefore observed in \textit{all}
stars later than $\sim$B0.5Ia in our sample, and it is possible that the 
early- to mid-B supergiants in our sample all lie within a region of 
pulsational instability that extends to the cool hypergiants \citep{mr72, 
schaller90}. Unless the stars are short-period binaries at favourable 
inclination (as with \object{W43a}, which displays clear signs of 
\textit{both} orbital and pulsational variations; see Figure~\ref{fig:w43a}) 
we cannot conclusively detect binarity with our short baseline of observations, although 
the longer baseline survey will be sensitive to long-term trends in these 
objects.  

\section{Discussion and Conclusions}\label{sec:discussion}

\subsection{Binary detections}

In this work we present first-epoch results from a VLT/FLAMES radial velocity
survey designed to identify high-mass binaries in the starburst cluster 
Westerlund~1. As expected, large RV changes are detected in the short-period 
eclipsing binary \object{W13} \citep{bonanos07} which displays variable emission 
from a B0.5Ia$^{+}$/WNVL emission line primary and He~I and infilled 
Paschen-series absorption in an early-B secondary. As such, it appears to 
represent the immediate evolutionary precursor to the binary-rich Wolf-Rayet 
population in Wd1, such as the eclipsing binary \object{B/WR77o} \citep{bonanos07}.
The WC9d star \object{W239} (F/WR77n) also shows significant RV 
changes compatible with a short-period WR+OB CWB viewed at moderate inclination: the
OB secondary is already strongly implied by the presence of hot dust 
\citep{crowther06}, hard X-ray emission (also seen in \object{W13}; \citealt{clark08}) 
and non-thermal radio emission \citep{dougherty}. 
In addition we identify \object{W43a} as a new short-period binary with a 
massive, so-far unresolved companion. \object{W234} and \object{W3003} are 
probable binaries, and \object{W2a} is a strong candidate binary. Both
\object{W2a} and \object{W234} appear to be short-period binaries, with
\object{W234}, at spectral type O9.5Ib, earlier than the apparently 
pulsationally-variable early-B supergiants discussed in Section~\ref{sec:earlyB}. 
Finally, the measured RVs for \object{W3003} are consistent with limited sampling
of a short- or longer-period binary. Future observations will resolve the nature
of these systems. 

An second important group of objects in the context of our RV survey are 
\object{W6a}, \object{W15}, \object{W17} (included in the \textit{faint} lists) and 
\object{W30a}, which have all been identified as strong binary candidates in 
previous studies of Wd1. \object{W30a} is the strongest X-ray source 
in Wd1 after the magnetar \object{CXOU J164710.2-455216} \citep{clark08} and a 
factor of $\sim$10 more luminous than would be expected from a single 
star\footnote{The optical spectrum of \object{W30a} is also indicative
of an interacting binary \citep{clark08}.}, while \object{W6a} is a periodic variable 
\citep{bonanos07} that is also a hard X-ray source. 
Both display almost featureless I-band spectra with very broad, shallow  
Paschen-series absorption lines typical of spectroscopic binaries (see
also \citealt{neg09a}) and broad, variable H$\alpha$ emission \citep{clark08}. 
\object{W15} and \object{W17} are both detected at radio \citep{dougherty} 
wavelengths: the spectral index of \object{W15} suggests composite 
stellar+non-thermal emission\footnote{Purely thermal emission from W15 would 
imply a very high mass loss rate inconsistent with the observed spectral 
type.}, while the index of \object{W17} is unambiguously non-thermal. The 
radio observations therefore suggest that both are CWBs, 
although they are only weak detections at X-ray wavelengths \citep{clark08}. 
None of these objects show significant RV variations in our 
data. In the case of \object{W6} and \object{W30a} this is unsurprising, 
as the broad, weak I-band spectral features are insensitive to changes in 
RV due to the difficulty in determining the line centre, although at 
favourable inclination the large RV changes expected of a short-period, high-mass 
binary would still be apparent. In contrast, \object{W15} and \object{W17} display well-defined Paschen-series 
absorption lines but no significant variations in RV. However, 
in these cases it is likely that wider orbits are required in order for 
non-thermal radio emission to be detectable, and the relatively short baseline 
of our first observations remains insensitive to longer-period, radio-strong 
binaries. 

Radial velocity surveys are also expected to be incomplete, as a subset of objects 
will be viewed at unsuitable orbital inclinations while others do not display 
spectra that allow accurate RV measurement. There is also an element of chance 
in conclusively detecting binarity from a limited number of RV measurements, as 
is apparent from the observations of \object{W13} and \object{W43a}: only one 
of five observations \object{W43a} unambiguously identifies it as binary, 
while by chance three observations of \object{W13} fall at almost exactly the 
same orbital phase. However, while the short-period binary \object{W13} was 
detected photometrically \citep{bonanos07} and was also as a strong binary 
candidate from spectroscopic \citep{neg09a} and X-ray \citep{clark08} surveys, 
neither \object{W43a} or the probable/candidate binaries \object{W2a}, 
\object{W234} and \object{W3003} have been identified in previous work. 
Therefore, although the principle aim of our VLT/FLAMES survey is to build a 
dataset that is sensitive to long-period binaries, there is also great value 
to close-spaced RV measurements (e.g. the pair from 18/07/2008, MJD=54665.0 and 
24/07/2008, MJD=54671.1, separated by six days) that effectively detect 
short-period binaries that are not apparent from photometry or single-epoch 
spectroscopy. 

\subsection{Pulsations and the cluster velocity dispersion}

The cool hypergiants \object{W243} and \object{W265} are both 
undergoing photospheric pulsations, with \object{W265} showing strong 
spectroscopic similarities with the non-radially pulsating yellow hypergiant 
\object{$\rho$ Cas} \citep{lobel03}. Chromospheric activity is also apparent 
from core emission in the near-IR Ca~II triplet. We also find significant RV 
variations of $\sim$15--25kms$^{-1}$ in five early-B targets: while these could
be orbital, in \textit{all} targets later than spectral type $\sim$B1Ia these RV
changes are accompanied by changes in absorption line profile suggestive of 
photospheric pulsations. The two-month baseline of our initial data offers 
prevents us from distinguishing a limited sampling of a long-period orbit
from photospheric pulsations in the early-B supergiants, and
further observations are required to break this degeneracy. 
However, regardless of their origin, these variations imply that care must be 
taken in deriving a stellar velocity dispersion for Wd1 from a single epoch of 
data. \cite{bosch09} find that the binary population leads to a significant 
overestimation of the velocity dispersion for \object{NGC~2070} unless binaries
are removed from the sample, and the pulsational variations reported here may 
have a similar effect on infra-red determinations of the velocity dispersion 
for Wd1 (e.g. \citealt{mengel08}). We note that separate VLT/FLAMES observations 
of the F2Ia$^{+}$ YHG \object{W4} also reveal apparent pulsational variations in 
the Fe~I lines \citep{clark09} and consider it likely that photospheric pulsations 
are a general characteristic of the transitional B-type supergiant and cool hypergiant 
populations of Wd1.

\subsection{Binary fraction}

Despite the expected limitations of our short-baseline initial dataset, the 
results presented here support the high binary fraction implied from the 
earlier photometric and multiwavelength observations of Wd1. Considering just 
the sample presented here, RV measurements show \object{W13}, \object{W239} 
and \object{W43a} are binaries with \object{W234}, \object{W3003} and \object{W2a} 
likely candidates. Radio and X-ray measurements strongly suggest a further 
four targets (\object{W6}, \object{W15}, \object{W17} and \object{W30a}) 
\textit{are} binaries but RV variations are undetected - and, in two cases, 
potentially undetectable - in our VLT/FLAMES data. Any signature of orbital 
motion in the two cool hypergiants is masked by photospheric pulsations, but 
the LBV \object{W243} requires a hot companion to produce the observed He~I 
and Ly$\alpha$-pumped Fe~II and O~I emission lines that are incompatible with 
the cool-phase LBV primary \citep{ritchie09}. The YHG \object{W265} is also 
plausibly binary, being associated with a compact, resolved radio source 
\citep{dougherty} which could not be ionized by the YHG itself. While the 
source of ionizing photons could simply be the OB supergiant population of 
Wd1, the YHG \object{HR 8752} is also detected at radio wavelengths as a 
result of the wind being ionized by a B1V companion \citep{stickland78, 
piters88}. The hot companion also explains the observed [N~II] emission in 
the spectrum of \object{HR 8752}; notably [N~II]~$\lambda\lambda$6548, 6583 
emission is visible in VLT/FLAMES LR6-mode spectra of \object{W265} (but
not in the other YHGs; \citealt{clark09}) and [N~II]~$\lambda$6583 emission 
was also reported by \cite{w87}. 

Therefore even a conservative estimate that includes just the robust RV 
detections\footnote{\object{W13}, \object{W43a}, \object{W239} and \object{W3003}.}
and the two objects that lack RV detections but have binarity strongly implied by other
observations\footnote{\object{W30a}, a strong X-ray and spectroscopic binary candidate,
\citep{clark08}, and \object{W243}, which displays an emission line spectrum 
incompatible with an isolated cool hypergiant, \citep{ritchie09}.}
suggests a binary fraction of 30\% (6/20)
amongst our initial \textit{bright} sample of supergiant stars in Wd1, increasing to $\ge$40\%
if other candidate binaries are included\footnote{\object{W6a}, \object{W15} and \object{W17}
are not included in these percentages, as they are specifically selected from the \textit{faint} lists
as objects we believe to be binary but do not show RV changes in our initial data.}. Both percentages are 
likely to increase as more data become available. If the more evolved 
WR/WNVL binaries and transitional hypergiants are excluded, then we find $\sim$19\% (3/16)  
of the OB supergiants in our sample are strong binary detections, rising
to $\sim$30\% (5/16) if candidates are included. Robust estimates of the binary
fraction amongst the OB supergiant population of Wd1 will become available once
the full FLAMES dataset is analysed in Paper II of this series. However, these values
are broadly consistent with the high binary fraction in the Wolf-Rayet population 
\citep{crowther06}; future observations and Monte-Carlo simulations planned
for Paper III will reveal if the binary fraction of the two populations are equivalent. 
 
\begin{acknowledgements}

JSC gratefully acknowledges the support of an RCUK fellowship.  IN has been 
funded by grants AYA2008-06166-C03-03 and Consolider-GTC CSD-2006-00070 from 
the Spanish Ministerio de Ciencia e Innovaci\'on (MICINN). We thank
an anonymous referee for detailed and constructive comments.

\end{acknowledgements}

\appendix

\section{Radial Velocity measurements}
\begin{table*}
\caption{Error-weighted radial velocity measurements for the targets listed in Table~\ref{tab:targets}.$^{a, b}$}
\label{tab:radial}
\begin{center}
\begin{tabular}{lc|l|c|ccccccc|l}
           &      &               & Range$^{d}$    & \multicolumn{7}{c|}{Modified Julian Date}             & \\
ID         &Epochs& Line(s)$^{c}$ & (kms$^{-1}$)   & 54637 & 54646 & 54665 & 54671 & 54692 & 54695 & 54713 & Notes \\
\hline
\hline
&&&&&&&&&&&\\
W2a        &  5   & Pa11-16, He~I &$-51.0$ -- $-14.1$ & --      & $-49.3$ & $-56.2$ & $-33.9$ & $-54.1$ & --      & $-18.6$ & Cand. binary\\ 
W8b        &  5   & Pa11-16       &$-61.6$ -- $-40.3$ & --      & $-61.6$ & $-49.9$ & $-40.3$ & $-47.4$ & --      & $-50.7$ & \\ 
W21        &  5   & Pa11-14       &$-57.4$ -- $-31.1$ & --      & $-57.4$ & $-35.0$ & $-43.1$ & $-55.1$ & --      & $-31.1$ & \\ 
W23a       &  5   & Pa11-16       &$-62.4$ -- $-42.9$ & --      & $-47.9$ & $-55.6$ & $-56.6$ & $-42.9$ & --      & $-62.4$ & \\ 
W43a       &  5   & Pa11-14       &$-137.9$ -- $+8.1$ & --      &$-109.7$ &$-123.1$ & $+8.1$  & $-38.8$ & --      &$-137.9$ & Binary\\
W55$^{e}$  &  7   & Pa11-14       &                   & $-43.1$ & $-49.9$ & $-55.4$ & $-45.4$ & $-43.8$ & $-53.9$ & $-55.5$ & \\
W60$^{e}$  &  5   & Pa11-14       &                   & --      & $-40.5$ & $-48.3$ & $-45.8$ & $-48.2$ & --      & $-37.7$ & \\
W71        &  7   & Pa11-16, He~I &$-61.9$ -- $-38.7$ & $-57.3$ & $-52.9$ & $-39.4$ & $-47.3$ & $-44.8$ & $-46.4$ & $-62.4$ & \\
W78        &  5   & Pa11-15       &                   & --      & $-46.7$ & $-46.4$ & $-41.3$ & $-39.9$ & --      & $-49.5$ & Puls. BSG \\
W84        &  5   & Pa11-13       &                   & --      & $-53.8$ & $-59.0$ & $-54.9$ & $-54.6$ & --      & $-55.8$ & \\                                 
W234$^{e}$ &  7   & Pa11-13       &$-67.0$ -- $-30.8$ & $-67.0$ & $-66.9$ & $-66.2$ & $-30.8$ & $-46.4$ & $-56.9$ & $-32.7$ & Prob. binary \\                               
W239$^{f}$ &  5   & C~III~$\lambda$8664 & $-81.6$ -- $+14.1$ & -- & $-26.9$ & $-10.7$ & $+14.1$ & $-49.7$ & --    & $-81.6$ & Wolf-Rayet\\
W243$^{g}$ &  5   & N~I (1 \& 8)  &$-28.8$ -- $-20.4$ & --      & $-22.7$ & $-20.4$ & $-27.7$ & $-28.3$ & --      & $-28.8$ & Puls. LBV \\
W265$^{g}$ &  5   & N~I (1 \& 8)  & N/A$^{h}$         & --      & $-47.6$ & $-36.8$ & $-36.0$ & $-37.9$ & --      & $-49.8$ & Puls. YHG \\
W373$^{e}$ &  5   & Pa11-14       &                   & --      & $-50.8$ & $-50.9$ & $-45.8$ & $-53.8$ & --      & $-52.9$ & \\
W3002$^{e}$&  5   & Pa11-14       &                   & --      & $-36.1$ & $-38.6$ & $-47.7$ & $-39.4$ & --      & $-43.0$ & \\
W3003$^{e}$&  5   & Pa11-14       &$-54.6$ -- $-4.4$  & --      & $-4.4$  & $-43.7$ & $-54.6$ & $-22.0$ & --      & $-10.6$ & Prob. binary\\
W3004$^{e}$&  5   & Pa11-14       &$-62.3$ -- $-35.5$ & --      & $-47.9$ & $-62.3$ & $-54.3$ & $-35.5$ & --      & $-46.0$ & \\
W3005$^{e}$&  5   & Pa11-13       &                   & --      & $-43.9$ & $-46.9$ & $-52.4$ & $-51.6$ & --      & $-49.9$ & \\
\hline
\end{tabular}
\end{center}
$^{a}$Errors are not listed, but are $\le$4kms$^{-1}$ for the OB supergiants, $\le$3kms$^{-1}$ 
for the two cool hypergiants and $\lsim$10kms$^{-1}$ for the Wolf-Rayet \object{W239}.\\
$^{b}$The broad, weak Paschen-series absorption lines in \object{W30a} precluded accurate RV 
measurement, while RV measurements for the eclipsing binary \object{W13} discussed in Section~\ref{sec:w13}
will appear in a forthcoming paper.\\
$^{c}$The line(s) used for RV measurement.\\
$^{d}$The range of measured radial velocities is given for targets in which statistically-significant variations are detected.\\
$^{e}$ Displays a distorted Pa-16 line due to blending with C~III$\lambda$8500 and a weak Pa-15 line that yield discrepant radial velocities 
compared to the other Paschen-series lines.\\
$^{f}$ The C~III~$\lambda$8500 and C~III~$\lambda$8664 lines seen in the spectrum of the Wolf-Rayet \object{W239} give 
discrepant RV measurements ($\sim$25kms$^{-1}$; see discussion in Section~\ref{sec:w239}). The value for 
C~III~$\lambda$8664 is given here. \\
$^{g}$ The cool hypergiants \object{W243} and \object{W265} display rich absorption-line spectra, with a large number of potential lines
suitable for RV measurement. The values given here are the average of nine of the strongest unblended N~I absorption lines from
multiplets~1 and~8 \citep{moore}. Note that in both objects species-to-species variation in RV is seen, with clear 
evidence of velocity stratification in the photosphere of \object{W265} (see discussion in text). \\
$^{h}$ The range of observed radial velocities in \object{W265} varies depending on the absorption line used to make the measurement. 
See Section~\ref{sec:hypergiants}. \\
\end{table*}

\end{document}